\documentclass[prb,amsmath,amssymb,graphicx,reprint]{revtex4-1}
\usepackage{graphicx}
\usepackage{dcolumn}
\usepackage{bm}
\usepackage{booktabs}
\usepackage{tabularx}
\newcommand{\p}{$\%$}
\newcommand{\pat}{{at.}\%~}

\newcommand{\gt}{$\gamma^{\prime\prime\prime}$}

\newcommand{\gs}{$\gamma^{\prime}$}
\newcommand{\pn}{$\mathrm{R{_{N_2}}}$}

\newcommand{\GMM}{$\alpha^{\prime\prime}\mathrm{-Fe_{16}N_{2}}$}
\newcommand{\eFezN}{$\varepsilon-\mathrm{Fe_{3-z}N}$}
\newcommand{\jFeN}{$\zeta-\mathrm{Fe_{2}N}$}

\begin{document}
\title{Effect of Al doping on phase formation and thermal stability of iron nitride thin films}

\author{Akhil Tayal$^a$, Mukul Gupta$^a$}\email{mgupta@csr.res.in/dr.mukul.gupta@gmail.com}\author{Nidhi Pandey$^a$, Ajay Gupta$^b$, M. Horisberger$^c$, J. Stahn$^d$}
\address{$^a$UGC-DAE Consortium for Scientific Research, University Campus, Khandwa Road, Indore 452 001,India}
\address{$^b$Amity Center for Spintronic Materials, Amity University, Sector 125, Noida-201 303}
\address{$^c$Laboratory for Developments and Methods, Paul Scherrer Institut, CH-5232 Villigen PSI, Switzerland}
\address{$^d$Laboratory for Neutron Scattering and Imaging, Paul Scherrer Institut, CH-5232 Villigen PSI, Switzerland}

\begin{abstract}

In the present work, we systematically studied the effect of Al
doping on the phase formation of iron nitride (Fe-N) thin films.
Fe-N thin films with different concentration of Al (Al=0, 2, 3, 6,
and 12 at.\%) were deposited using dc magnetron sputtering by
varying the nitrogen partial pressure between 0 to 100\%. The
structural and magnetic properties of the films were studied using
X-ray diffraction and polarized neutron reflectivity. It was
observed that at the lowest doping level (2 at.\% of Al), nitrogen
rich non-magnetic Fe-N phase gets formed at a lower nitrogen
partial pressure as compared to the un-doped sample.
Interestingly, we observed that as Al doping is increased beyond
3at.\%, nitrogen rich non-magnetic Fe-N phase appears at higher
nitrogen partial pressure as compared to un-doped sample. The
thermal stability of films were also investigated. Un-doped Fe-N
films deposited at 10\% nitrogen partial pressure possess poor
thermal stability. Doping of Al at 2at.\% improves it marginally,
whereas, for 3, 6 and 12at.\% Al doping, it shows significant
improvement. The obtained results have been explained in terms of
thermodynamics of Fe-N and Al-N.

\end{abstract}

\date{\today}
\maketitle

\section{Introduction}
\label{1} Iron nitride compounds are known to exist in varieties
of phases having distinct crystal structure and magnetic
properties~\cite{Gupta_JAC01,Gupta:PRB05,Liu:JAP:2003,Rissanen_JAC98,Schaaf.PMS.2002,Jiang:1995,PRB:Ranu:2006,JAC:Peng:1997,Easton:TSF05,ML:2012,Vacuum:Li:2009,Hinomura.PB.1997,ASS:Wang:2003}.
Different Fe-N phases can be formed just by increasing the
nitrogen concentration in Fe. With the increasing nitrogen
concentration, Fe-N phases that get formed are:
nanocrystalline/amorphous $\alpha$-Fe-N (N\pat=0-11), \GMM
(N\pat=11.4), \gs-Fe$_4$N (N\pat$\sim$20), \eFezN
(0$\leq$z$\leq$1, N\pat=25-33), $\zeta$-Fe$_2$N (N\pat$\sim$33),
\gt-FeN (N\pat$\sim$50). This essentially covers the whole phase
diagram of Fe-N system~\cite{PD:FeN:2010}. These compounds have
various applications such as in tribological coatings, magnetic
read-write heads, memory devices
etc.~\cite{Schaaf.PMS.2002,PRB:Ranu:2006,Navio.PRB08} However,
thermal stability of Fe-N compounds is poor due to weak Fe-N
bonding and since the heat of formation ($\Delta H_f^{\circ}$) for
Fe-N is high (compared to other transition metal nitrides e.g.
Al-N, Ti-N etc.), Fe-N compounds are invariably less stable
~\cite{Gupta_PRB02,Gupta:AM:2009,Ding:IEEE:2006,gupta:JAP2011,Chechenin:JPCM:2003,Tessier_SSS00,Kopcewicz:JAP:1995}.
To improve the thermal stability of Fe-N system, it was proposed
that if a few atomic percentage of a third element X (X=Al, Ti,
Zr, Ta etc.), which has low $\Delta H_f^{\circ}$ for its nitrides
and high affinity nitrogen, is added in the Fe-N system, then it
can thermally stabilize the iron nitride
compounds~\cite{Chechenin:JPCM:2003,Viala:JAP:1996,Wang:JPCM:1997,Varga:JAP:1998,
Liu:APL:2000,Chezan:PSSA:2002,Rantschler:JAP:2003,Liu:JAP:2003,Kazmin:TPL:2005,
Das:PRB:2007,Sangita:PRB:2008,FengXu:JAP:2011,RG:JAP12}. Very
recent study elucidates that enhancement of thermal stability
actually results from the suppression of Fe
self-diffusion~\cite{PRB:AT:2014,JAP:AT}. In addition, it was
observed that Al doping is most efficient (compared to other
dopants e.g. Zr, Ti) for the enhancement of thermal
stability~\cite{PRB:AT:2014,JAP:AT}. However, there were no
reports mentioning the optimum amount of Al that would be require
for the enhancement of the thermal stability. Moreover, as all the
proposed dopants are non-magnetic the amount of dopants is very
crucial, since it may alter the ingenious magnetic properties of
Fe-N thin films. Therefore, it is required that an optimum doping
level must be known so that it would not affect the desired
properties of Fe-N thin films.

In the present work, we have addressed these issues by
systematically studying the formation of different iron nitride
phases by varying the doping level of Al. Structural properties of
the films were investigated using X-ray diffraction (XRD).
Polarized neutron reflectivity (PNR) was used to measure the
magnetic properties of the deposited samples. We observed that at
low doping of Al(2\,at.\%), nitrogen incorporation in the Fe-N
system gets enhanced as compared to un-doped sample.
Interestingly, we observed that with increasing the doping level
of Al from 3 to 12\,at.\%, nitrogen incorporation in iron nitride
system gets retarded. As expected, the thermal stability of
un-doped samples is poor. With 2at.\% of Al, it only improves
marginally. Whereas, Al doping at 3, 6 and 12at.\% shows
significant enhancement in the thermal stability. The obtained
results explained on the basis of interaction of Al with N.

\section{Experimental}
\label{2}

Iron nitride films were deposited using a dc magnetron sputtering
(dcMS) technique. Pure Fe and [Fe+Al] composite targets were
sputtered using a mixture of Ar and N$_2$ gases. Nitrogen
concentration in the films was varied by varying the nitrogen
partial pressure defined as: \pn =
$\mathrm{P{_{N_2}}}$/($\mathrm{P{_{Ar}}}$+$\mathrm{P{_{N_2}}}$),
(where $\mathrm{P{_{Ar}}}$ is Ar gas flow and $\mathrm{P{_{N_2}}}$
is N$_2$ gas flow) between 0 to 100\%. Total gas flow was kept
constant at 10\,sccm. Before deposition a base pressure of
1$\times10^{-5}$\,Pa was achieved, during the deposition chamber
pressure was maintained at 0.4\,Pa. Total five sets of samples
were prepared by varying the \pn~between 0 to 100\% at Al doping
of 0, 2, 3, 6, and 12 at.\%. Concentration of Al in the samples
was measured using energy dispersive X-ray analysis, typical error
bars in the measurement are of the order of $\pm$\,0.5. All the
samples at one particular concentration of Al were prepared
simultaneously. The structural characterization of the samples was
performed using a standard XRD system (Bruker D8-Advance) equipped
with Cu-K$\alpha$ x-ray source and a one dimensional position
sensitive detector based on silicon strip technology (Bruker
LynxEye) in $\theta-2\theta$ geometry. PNR measurements were
carried out at AMOR reflectometer at SINQ-PSI Switzerland. To
saturate the sample magnetically a magnetic field of 0.5\,T was
applied during the PNR measurements.

\section{Results}
\label{RsandDis}

\subsection{Phase evolution in Fe-Al-N thin films}
\label{PhaseEvol}

\begin{figure} \center
\includegraphics [width=80mm,height=90mm] {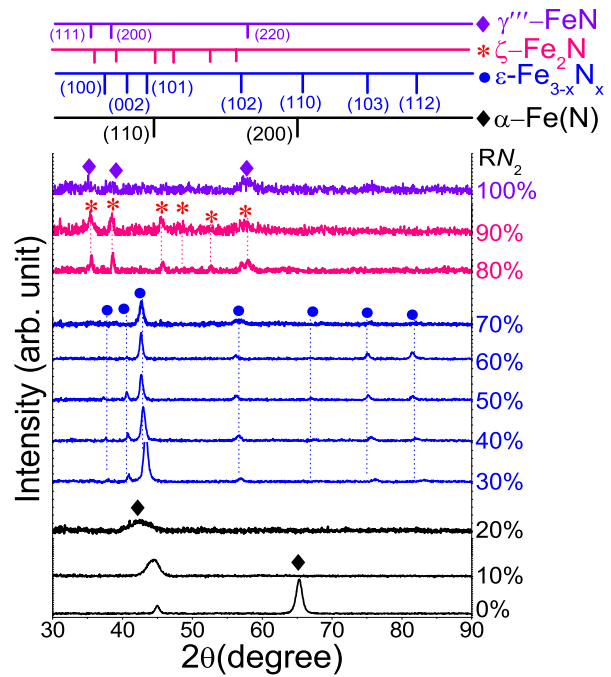} \caption{\label{fig:XRDFe_N} XRD
patterns of un-doped iron nitride thin films deposited at
different \pn.}

\end{figure}

As such formation of different Fe-N phases with increasing \pn~has
been well-studied~\cite{RG:JAP12,TSF:Tayal:13}. Here to make an
unambiguous comparison of phase formation for various Al doping,
un-doped samples were also prepared under identical deposition
conditions. Figure~\ref{fig:XRDFe_N} shows XRD patterns of
un-doped Fe-N thin films prepared for different \pn. The observed
Fe-N phases with increasing \pn~and respective crystallite size
are tabulated in the table~\ref{tab:xrdphaseGS}. Crystallite size
in the samples was calculated using Scherrer
formula~\cite{Scherrer1918,Cullity_XRD}. From the XRD pattern, it
can be seen that up to \pn=20\p~nanocrystalline/amorphous
$\alpha$-Fe(N) phase with $bcc$ structure can be seen. For
\pn~between 30 and 70\p, \eFezN (0$\leq$z$\leq$1) phase was
observed. The $\varepsilon$ phase exist in a wide range of
nitrogen composition varying from 0$\leq$z$\leq$1. When
0$\leq$z$\leq$0.6 the phase is ferromagnetic at room temperature,
whereas above this it becomes paramagnetic at room
temperature~\cite{JPC:chen:FexN:1983}. For \pn=80 and 90\p,
\jFeN~phase having orthorhombic structure was observed while for
\pn=100\p, \gt-FeN phase can be seen. These results are
consistence with previous reports~\cite{RG:JAP12,TSF:Tayal:13}.

\begin{table*}
\caption{\label{tab:xrdphaseGS} Crystallite size (CS in nm) and
phases formed in Fe-Al-N thin films samples with Al doping of 0,
2, 3, 6 and 12\pat~deposited by varying the nitrogen partial
pressures between \pn =0 - 100\p. Here $am$ denotes amorphous
phase. The error bar in measurement of crystallite size is about
$\pm$0.5\,nm. Indexation of various Fe-N phases was done following
the JCPDS database and XRD data reported in the literature, for
$\alpha$-Fe-N (JCPDS No. 85-1410), $\gamma^\prime$-Fe$_{4}$N
(JCPDS No. 86-0231), \eFezN~(JCPDS No. 03-0925 and ref.
~\cite{SchaafHypInt95}), \jFeN~(JCPDS No. 86-1025 and
ref.~\cite{Cai_2000,JAC_Rechenbach_96}) and \gt-FeN (JCPDS No.
88-2153 and ref.~\cite{Jouanny2010TSF}).} \vspace{5mm}
\begin{tabular}{|c|cc|cc|cc|cc|cc|}
\hline \pn &\multicolumn{2}{|c|}{0\pat~Al}&\multicolumn{2}{|c|}{2\pat~Al}&\multicolumn{2}{|c|}{3\pat~Al}&\multicolumn{2}{|c|}{6\pat~Al}&\multicolumn{2}{|c|}{12\pat~Al}\\
\cline{2-3} \cline{4-5} \cline{6-7} \cline{8-9} \cline{10-11}
(\p)&Phase&CS&Phase&CS&Phase&CS&Phase&CS&Phase&CS\\ \hline
0&$\alpha$&15&$\alpha$&14.9&$\alpha$&23.3&$\alpha$&24.1&$\alpha$&19.6 \\
10&$\alpha$&4.6&$\alpha$&7.7&$\alpha$&12.9&$\alpha$&14.7&$\alpha$&15.9 \\
20&$\alpha$&$am$&$\alpha$&$am$&$\alpha$&7.4&$\alpha$&9.6&$\alpha$&11.0 \\
30&$\varepsilon$&10.6&$\varepsilon$&13.9&$\alpha$&3&$\alpha$&$am$&$\alpha$&$am$ \\
40&$\varepsilon$&14.9&$\varepsilon$&20.7&$\varepsilon$&15.8&$\alpha$&$am$&$\alpha$&$am$ \\
50&$\varepsilon$&21.2&$\varepsilon$&29.5&$\varepsilon$&24.8&$\varepsilon$&14.9&$\alpha$&$am$ \\
60&$\varepsilon$&17.5&\gt&10.2&$\varepsilon$&24.8&$\varepsilon$&15.4&$\varepsilon$&4.0 \\
70&$\varepsilon$&14.7&\gt&6.0&$\varepsilon$&19&$\varepsilon$&14.4&$\varepsilon$&4.0 \\
80&$\zeta$&19.7&\gt&3.3&\gt&7.4&($\varepsilon$+\gt)&10.5&$\varepsilon$&$am$ \\
90&$\zeta$&7.6&\gt&3.0&\gt&5.4&\gt&4.4&\gt&2.5 \\
100&\gt&5.0&\gt&3.2&\gt&3.8&\gt&3.4&\gt&2.4 \\
\hline
\end{tabular}
\end{table*}

\begin{figure*} \center
\includegraphics [width=100mm,height=90mm] {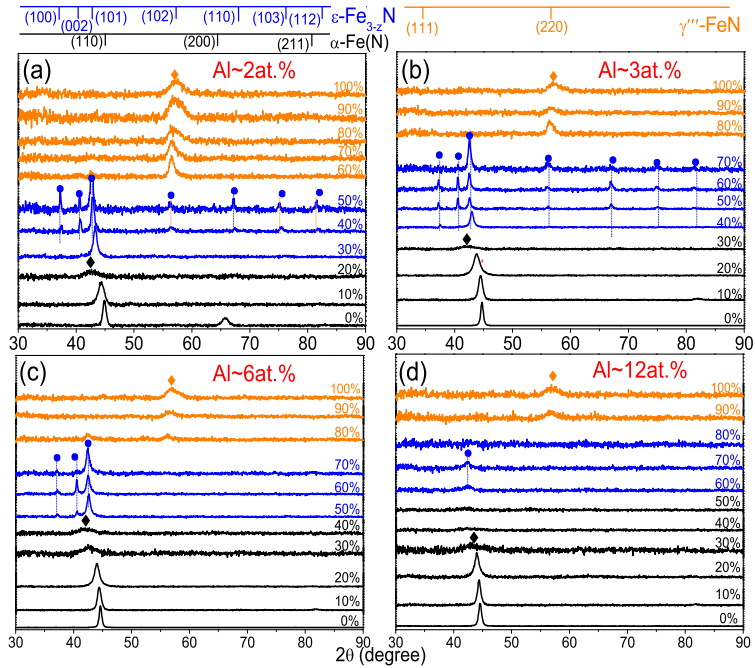}
\caption{\label{fig:XRDFeAlN} XRD patterns of Fe-N thin films
deposited at different \pn~with Al doping of 2at.\%(a), 3at.\%(b),
6at.\%(c) and 12at.\%(d).} \vspace{-1mm}
\end{figure*}

Now the effect of Al doping on the phase formation can be
compared. Figure~\ref{fig:XRDFeAlN} shows XRD patterns of samples
prepared with different amount of Al doping at 2\%(a), 3\%(b),
6\%(c), and 12\%(d). It can be seen that as compared to the
un-doped samples, the doping of Al has altered the order of phase
formation while the phases that occur are similar to the un-doped
case. The observed Fe-N phases and their crystallite size are
given in the table~\ref{tab:xrdphaseGS}. With 2at.\% Al doping
(figure~\ref{fig:XRDFeAlN}(a)), up to \pn=20\% the phases formed
are similar as they occur in the un-doped sample. However, the
$\varepsilon$ phase can only be seen between \pn=30, 40 and 50\p.
Above this value of \pn, \gt~phase gets formed. No signature of
\jFeN~phase can be seen for samples prepared with Al doping.

We find that as the Al doping levels are increased, the value of
\pn~at which the $\varepsilon$ phase starts shifts to a higher
value. For 3at.\% it is 40\%, for 6at.\% it is 50\% and for
12at.\% it is 60\%. This also suggests that value of \pn~for which
nanocrystalline/amorphous $\alpha$-Fe(N) phase is seen gets
extended with an increase in Al doping. Other variations remain
almost similar as shown in figure~\ref{fig:XRDFeAlN}(a),(b) and
(c) for samples prepared with Al doping of 3,6 and 12at\%,
respectively. A general trend that emerges out is that with
successively increasing Al doping, formation of nitrogen rich
phases occurs at higher \pn.

\subsection{Study of magnetic properties using polarized neutron reflectivity}
\label{PNR}

\begin{figure*} \center
\includegraphics [width=100mm,height=80mm] {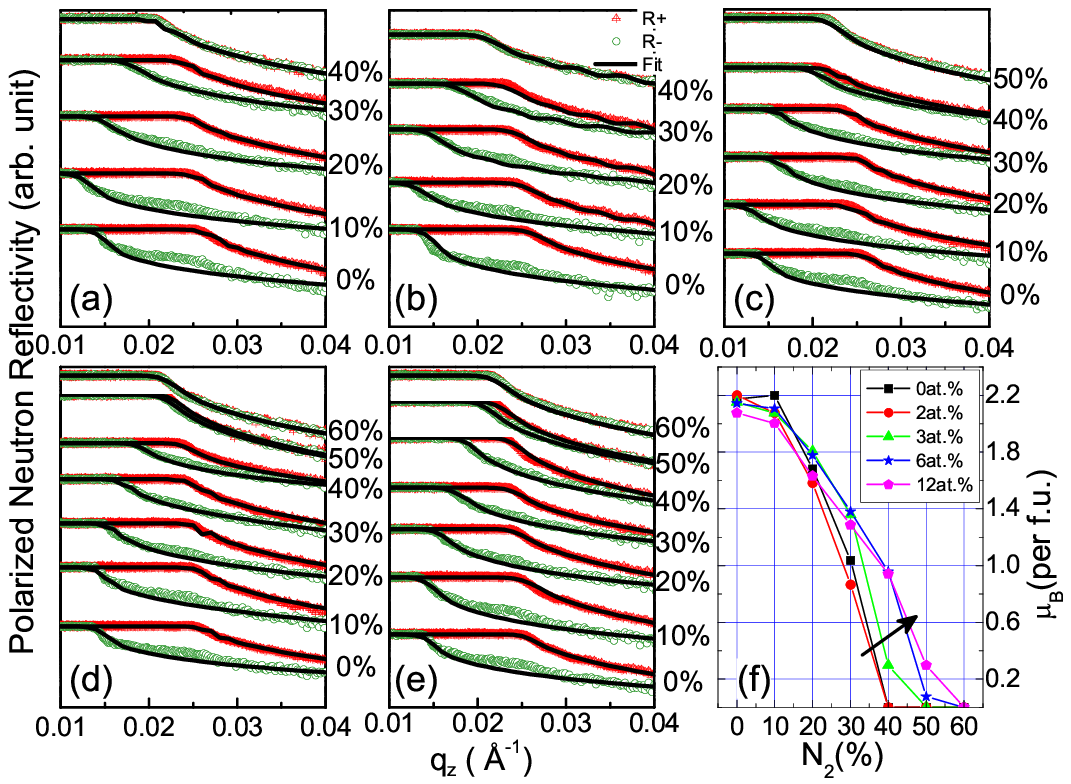}
\caption{\label{fig:PNRFeAl} PNR patterns of samples deposited at
different \pn~with Al doping of 0at.\%(a), 2at.\%(b), 3at.\%(c),
6at.\%(d), and 12at.\%(e). Variation of magnetic moment with
increasing \pn~obtained from fitting PNR patterns(f).}
\vspace{-1mm}
\end{figure*}

To investigate the implication of doping concentration on the
magnetic properties of deposited samples, we have performed PNR
measurements on all set of samples. PNR is a well known technique
for the measurement of magnetic moment in thin
films~\cite{Blundell_PRB92}. Magnetic moment obtained from this
technique as compared to conventional magnetization technique has
an unique advantage, that it is free from errors that are
generally occurred in measuring sample mass. PNR measurements on
all sets of samples were carried out up to the value of \pn, where
spin-up ($R^{+}$) and spin-down ($R^{-}$) reflectivities equals.
When $R^{+}=R^{-}$, it suggests that the sample has became non
ferromagnetic at the measured temperature.
Figure~\ref{fig:PNRFeAl} shows PNR patterns of samples deposited
at varying \pn~for Al=0at.\%(a), 2at.\%(b), 3at.\%(c), 6at.\%(d),
and 12at.\%(e).

\begin{figure} \center
\includegraphics [width=80mm,height=70mm] {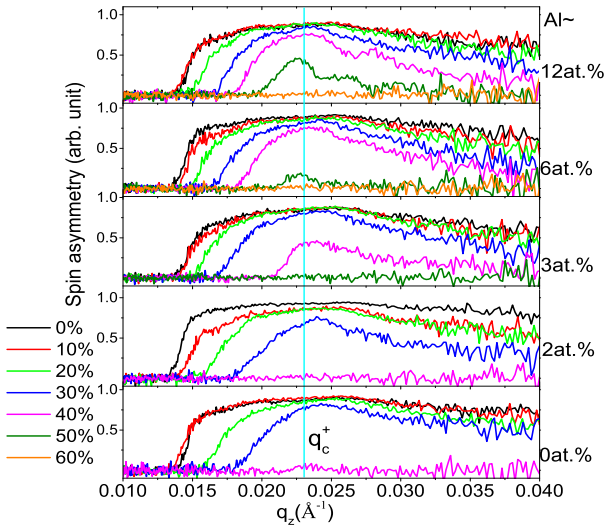}
\caption{\label{fig:SpinAsymm} Spin asymmetry obtained from PNR
patterns for samples deposited at different \pn~with Al doping of
0, 2, 3, 6 and 12at.\%.} \vspace{-1mm}
\end{figure}

Raw PNR data even without fitting can be used to get vital
information about the magnetic structure of the samples. Spin
asymmetry (SA= ($R^{+}-R^{-})/(R^{+}+R^{-}$)) can be used to
observe variation in its intensity which is proportional to
magnetic moment of the sample. Figure~\ref{fig:SpinAsymm} show a
plot of SA for samples deposited at varying \pn. If the relative
intensity at around q$_z^{+}$ is monitored, it gives qualitative
information regarding the variation in magnetic moment with
increasing \pn. In the un-doped sample, SA intensity at q$_z^{+}$
falls to zero for sample deposited at \pn=40\p. For 2at.\% Al, SA
intensity is zero for \pn=40\p~sample, however, for 30\p~sample,
SA intensity at q$_z^{+}$ is slightly reduced compared to un-doped
sample deposited at same \pn. This indicates that at 2at.\% Al
doping, nitrogen rich non-ferromagnetic Fe-N phase gets formed at
relatively less \pn. On contrary, at 3at.\% Al doping, SA
intensity is zero for sample deposited at \pn=50\p. With further
increasing Al concentration to 6 and 12 at.\%, SA intensity is
zero for 60\p~samples. However, on comparing these two set of
samples SA intensity falls more rapidly for 6at.\% Al doping as
compared to 12at.\% Al doping.

To get quantitative information about the variation of magnetic
moment in the samples, PNR data was fitted using a computer
program~\cite{SimulReflec}. Obtained values of the average
magnetic moment are plotted in figure~\ref{fig:PNRFeAl}(f). It can
be seen that with increasing \pn~magnetic moment falls to zero in
all samples. However, with doping the fall in magnetic moment get
altered. For low doping (2at.\%) moment fall more rapidly as
compared to un-doped samples. Whereas, with increasing Al doping
beyond 3at.\% fall in the value of magnetic moment continuously
shifts to a higher \pn. These results indicate that at low Al
doping (2at.\%) formation of nitrogen rich Fe-N phases occurs at
relatively lower \pn~as compared to un-doped samples. On contrary
to this, as doping concentration of Al increases, nitrogen rich
Fe-N phase get formed at relatively higher \pn~as compared to
successive lower Al doping. These results are consistence with the
XRD measurements discussed in section~\ref{PhaseEvol}.

\subsection{Investigation of thermal stability}
\label{TS}

\begin{figure*} \center
\includegraphics [width=100mm,height=80mm] {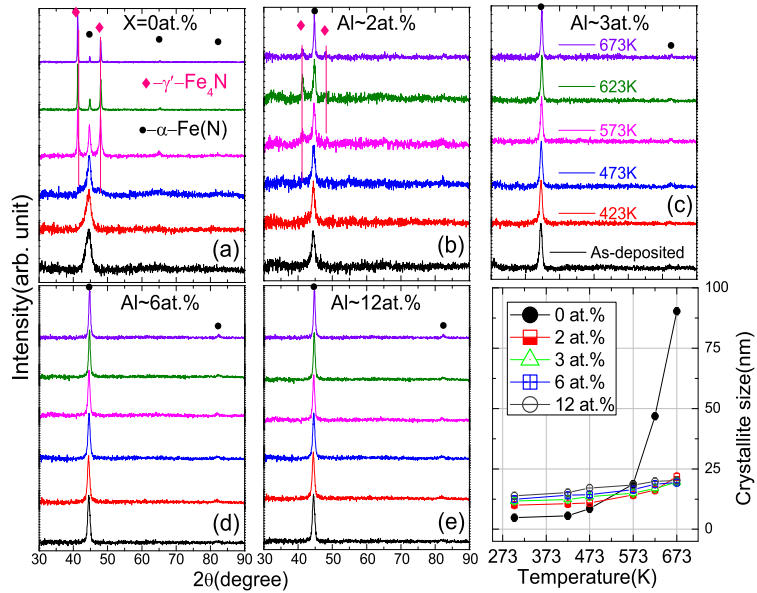}
\caption{\label{fig:XRDTS} XRD patterns of samples deposited at
\pn=10\% with Al doping of 0at.\%(a), 2at.\%(b), 3at.\%(c),
6at.\%(d), and 12at.\%(e) in the as-deposited state and after
annealing at various temperatures.}
\end{figure*}

To investigate the effect of Al concentration on the thermal
stability of Fe-N thin films, we studied Al doped and un-doped
samples after annealing them at different temperatures. We
selected five samples deposited at \pn=10\% for various Al doping.
At this value of \pn~the structure of all samples is similar
(nanocrystalline $\alpha$-Fe-N phases) and  samples are expected
to show soft magnetic properties~\cite{JAP:AT,SCT15_AT}. When N
atoms are incorporated in $bcc$-Fe, they occupy interstitial sites
within the Fe lattice. Since the radius of the interstitial site
is almost half of a N atom~\cite{Gupta:PRB05}, N incorporation
give rise to lattice strain. In this process strain energy
dominates over grain boundary energy that leads to reduction in
grain size. If the grain size in the films reduces below
ferromagnetic exchange length, in such condition, films displays
excellent soft magnetic properties as explained by the random
anisotropy model~\cite{Herzer_IEEE89,Loffler_PRB98}. Therefore,
such kind of films are important for device application and hence
their thermal stability is necessary to investigate.

To study the effect of Al doping on the thermal stability all
sample prepared at \pn=10\% were annealed in a in a vacuum furnace
for 2\,hours. To reduce any thermal gradient, all samples were
annealed together. After an annealing at a temperature, XRD
measurements were performed. Figure~\ref{fig:XRDTS} shows XRD
patterns of annealed samples at various temperatures. It can be
seen in figure~\ref{fig:XRDTS}(a) that in the un-doped sample
nanocrystalline $bcc$-Fe(N) phase remains stable only up to
423\,K. Above this temperature XRD peaks corresponding to
\gs~phase starts appearing that continuously grow in intensity
with an increase in annealing temperature. With 2at.\% Al doping,
(figure~\ref{fig:XRDTS}(b)) thermal stability of nanocrystalline
$bcc$-Fe(N) phase only improves marginally as small peaks
corresponding to \gs~phase starts to appear above 573\,K. In
contrast to it, the thermal stability of 3, 6 and 12 at.\% Al
doped sample, improves remarkably as no additional peaks
corresponding to any other Fe-N phase can be detected even after
annealing at 673\,K.

Although thermal stability improves significantly when Al doping
in Fe-N system exceeds 3at.\%, it is immensely important that the
grain growth should be avoided as annealing temperature increases.
For grain sizes below the ferromagnetic exchange length, soft
magnetic properties deteriorate as the grain size increase. The
grain sizes obtained from XRD data for annealed samples are shown
in figure~\ref{fig:XRDTS}(f). As expected in the un-doped sample
sudden grain growth was observed above 573\,K. Whereas with Al
doping grain size remains stable throughout the annealing
temperatures. The measurement of magnetic properties on similar
samples were done in earlier studies~\cite{RG:JAP12,TSF:Tayal:13}.
It was found that with doping soft magnetic properties of films
shows significant improvement.

\section{Discussion}

The obtained results can be understood from the interaction of N
atoms with Fe and Al. It is know that the $\Delta H_f^{\circ}$ for
aluminum nitrides is significantly smaller than that of iron
nitrides ($\Delta H_f^{\circ}$ for Al-N is typically -321\,
kJmol$^{-1}$; for Fe-N it is about -10\, kJmol$^{-1}$) and Al has
far more affinity for N as compared to Fe~\cite{PRB:AT:2014}. As N
is added in Fe, the magnetic moment falls due the formation of
non-magnetic covalent bonds between Fe and N. However, as
concentration of Al doping increases, the formation of
non-magnetic iron nitride shows a non-monotonic behavior with
\pn~as compared to un-doped samples. This can be understood from
the thermodynamics of Al-N. When Al is doped in Fe, it may either
get dissolved substitutionally or may get incorporated in the
grain boundary region. Al atoms present within the Fe lattice
enhance the N incorporation since $\Delta H_f^{\circ}$ is small
and high affinity for N. Enhanced N content increases the
interaction of Fe with N that results in the formation of
non-magnetic Fe-N phase at relatively smaller \pn~as compared to
un-doped sample. This phenomenon was indeed observed with 2at.\%
Al doping in the present study as well as in previous
reports~\cite{TSF:Tayal:13,RG:JAP12}. However with increasing Al
concentration beyond 3at.\% a reverse effect can be observed.
Higher doping level enhance the probability for Al atoms to be
present in the grain boundary region. During the deposition, N
atoms interact with these Al atoms to form Al-N. This behavior
shields a direct interaction between Fe and N atoms. Due to
reduced interaction between Fe and N formation of non-magnetic
Fe-N phases shifts to higher \pn~with increasing doping
concentration. The grain boundary precipitate layer in the form of
Al-N also act as a diffusion barrier layer that suppress Fe
self-diffusion which results in the observed enhancement in the
thermal stability of Fe-N films at 3, 6 and 12at.\% of Al.
Moreover no grain growth is observed at these doping level
attributed to suppression in atomic diffusion. From the obtained
results, it can be concluded that to get good soft magnetic
properties and remarkable thermal stability, optimum doping level
of Al in the Fe-N thin films must be about 3at.\%.

\section{Conclusion}
\label{4}

In conclusion, Fe-N films were deposited using dc-MS by varying
the \pn. Phase formation in the Fe-N system was compared for
different Al doping (2, 3, 6 and 12at.\%). At low Al doping
(2at.\%) nitrogen rich Fe-N phases gets formed at lesser \pn~as
compared to un-doped samples. On the contrary, successive increase
in Al doping levels shift the formation of non-ferromagnetic Fe-N
phase to higher \pn. Thermal stability of un-doped sample
deposited at \pn=10\% is found to be poor. With 2at.\% thermal
stability only improves marginally. Whereas, for 3, 6 and 12at.\%
thermal stability improves remarkably while the grain size remains
almost constant. We find that when doped sufficiently, Al atoms
are not only substitutionally dissolved but also present in the
grain boundary region. Formation of a diffusion barrier layer in
the grain boundary region prevents grain growth and leads to
remarkable thermal stability.

\section*{Acknowledgments}
A part of this work was performed at AMOR, Swiss Spallation
Neutron Source, Paul Scherrer Institute, Villigen, Switzerland. We
acknowledge Department of Science and Technology, New Delhi for
providing financial support to carry out PNR experiments. We would
like to acknowledge Dr.\,A.\,K.\,Sinha and Dr.\,V.\,Ganesan for
support and encouragement. One of the authors (AT) wants to
acknowledge CSIR, New Delhi for the senior research fellowship.


%

\end{document}